\documentclass[twocolumn, showpacs,prl,amsmath,amssymb,amsthm,superscriptaddress]{revtex4}

\usepackage{graphicx}
\usepackage{bm}

\begin{document}
\title{Electron spin decoherence in isotope-enriched silicon}
\author{Wayne M. \surname{Witzel} \email{wwitzel@sandia.gov}}
\affiliation{Sandia National Laboratories, Albuquerque, New Mexico 87185 USA}
\author{Malcolm S. \surname{Carroll}}
\affiliation{Sandia National Laboratories, Albuquerque, New Mexico 87185 USA}
\author{Andrea \surname{Morello}}
\affiliation{CQCT and School of
  Elec. Eng. \& Telecom., University of New South Wales, Sydney, Australia}
\author{{\L}ukasz \surname{Cywi{\'n}ski}}
\affiliation{Institute of Physics, Polish Academy of Sciences,
02-668 Warszawa, Poland}
\affiliation{University of Maryland, College Park, Maryland 20742-4111, USA}
\author{S. \surname{Das Sarma}}
\affiliation{University of Maryland, College Park, Maryland 20742-4111, USA}

\begin{abstract}
Silicon is promising for spin-based quantum computation because nuclear
spins, a source of magnetic noise, may be eliminated through isotopic
enrichment. Long spin decoherence times $T_2$ have been measured in
isotope-enriched silicon but come far short of the
$T_2 = 2 T_1$ limit.  The effect of nuclear spins on $T_2$ is well
established.  However, the effect of background electron spins from
ever present residual phosphorus impurities in silicon can
also produce significant decoherence.  We study spin decoherence decay
as a function of donor concentration, $^{29}$Si concentration, and
temperature using cluster expansion techniques specifically adapted to 
the problem of a sparse dipolarly coupled electron spin bath.  Our
results agree with the existing experimental spin echo data in Si:P
and establish the importance of background dopants as the ultimate
decoherence mechanism in isotope-enriched silicon.
\end{abstract}
\maketitle

Long electron spin decoherence times in silicon are of significant interest in
producing low-error rates for quantum computation.  Very long
spin echo decay times $T_2$ have been reported~\cite{tyryshkin60ms, Abe10, tyryshkin600ms} in isotope-enriched
silicon (i.e., reduced nuclear spin concentration).  
Resource requirements for quantum error correction are
significantly reduced as the qubit fidelity improves, which motivates
better understanding of the limits of $T_2$ even with isotope
enrichment. 
The ultimate decoherence time is theoretically limited by inelastic decay
mechanisms (spin-lattice relaxation) on a time scale of $T_1$.  
The increase of $T_2$ upon reducing the $^{29}$Si nuclear spin
concentration~\cite{tyryshkin60ms, Abe10} is now well understood~\cite{WitzelSiP, SemionLinkedCluster}.  
However, even the highest purity Si wafers contain traces of dopant
impurities, usually phosphorus, at levels $\sim 10^{12} -
10^{14}$~cm$^{-3}$.
Their electron spins are coupled by dipolar interactions,
causing fluctuations that induce qubit spin dephasing.
In this Letter, we develop the necessary theory to examine decoherence
of a central spin in a sparse bath of nuclear and electron spins.  We find
excellent agreement with
existing Si spin echo data showing that existing spin decoherence
measurements in Si may already be limited by the
coupling of the donor electron spin to the P donor spin bath rather
than the Si nuclear spin bath.  
As a consequence, further isotopic enrichment,
an extremely expensive procedure, may not provide any more advantage
in the eventual construction of a Si spin quantum computer.  In fact,
we find that in the presence of donor-induced spin decoherence, $T_2$ may
actually increase when some $^{29}$Si is present.

We study here  the central spin decoherence problem of a donor electron spin among spins of other donors and $^{29}$Si.  
Because of coupling among the spins, a particular donor electron spin will
experience fluctuations of its energy splitting in a
phenomenon known as spectral diffusion (SD).
$^{29}$Si-induced SD calculated using a cluster expansion technique~\cite{WitzelSiP}, well
approximated at the lowest order with independent contributions from
each pair~\cite{ShamPairApprox}, is in excellent agreement with
experiments for Si:P~\cite{tyryshkin60ms, Abe04, Abe10}
and Si:Bi~\cite{Morton10} donors.
With a firm foundation rooted in a precise quantum mechanical formulation, this was a significant advance over the long history of phenomenological, stochastic models~\cite{Stochastic, deSousaDonors, deSousaSD}.
  These previous techniques~\cite{WitzelSiP, ShamPairApprox}, however, are applicable to relatively dense and weakly-coupled spin baths and cannot accurately treat SD due to  randomly located donors in which the strength of interaction to the central spin is no different than between bath spins; neither can they handle very low concentrations of $^{29}$Si rigorously.
 A disjoint cluster approach was applied to the relatively sparse bath of carbon spins for the SD of nitrogen-vacancy defects in diamond~\cite{TaylorNV}.  Exact numerics were applied~\cite{SlavaDipolarBath} in the central spin decoherence problem of dilute dipolarly-coupled spins. 
  Our approach in this Letter is based upon the cluster correlation
  expansion (CCE)~\cite{YangCCE} that reformulates the cluster
  expansion technique~\cite{WitzelSiP} such that a large bath
  approximation is not necessary, making new regimes of the SD problem
  accessible.

We consider an ensemble of Si:P donor electron spins over varied donor
concentrations, $C_E$ (for electron), and $^{29}$Si concentrations,
$C_N$ (for nuclear).  We use parts per
million (ppm) of lattice sites for $C_N$.
We include dipolar and hyperfine interactions among spins.
Assuming a large applied
magnetic field ($100~\mbox{mT}$ is sufficient) in the
$z$ direction, we use an effective Hamiltonian in which Zeeman
energies are conserved
among the electron and nuclear spins independently, allowing only
flip-flop dynamics:
$\hat{\cal H} \!=\! \hat{H}_{E} + \hat{H}_{N} +
\hat{H}_{E-N}$, where
\begin{eqnarray}
\label{Eq:H_E}
\hat{H}_{E} &=& \sum_{i > j} \gamma_E^2 d({\bf R}_{i} - {\bf R}_j) [\hat{S}_i^+ \hat{S}_j^-
 + \hat{S}_i^- \hat{S}_j^+ - 4 \hat{S}_i^z \hat{S}_j^z],~~~~
\\
\label{Eq:H_N}
\hat{H}_{N} &=& \sum_{n > m} \gamma_N^2 d({\bf r}_{n} - {\bf r}_m)
[\hat{I}_n^+ \hat{I}_m^-
 + \hat{I}_n^- \hat{I}_m^+ - 4 \hat{I}_n^z \hat{I}_m^z],~~~~
\end{eqnarray}
with the dipolar interaction strength given by $d({\bf r}) \! = \! [1 - 3 (r_z/r)^2]/4 r^{3}$, and
 \begin{eqnarray}
\label{Eq:H_E-N}
\hat{H}_{E-N} &=& \sum_{i, n} \gamma_E \gamma_N h_i({\bf R}_{i} - {\bf
  r}_n) \hat{S}_{i}^{z} \hat{I}_{n}^{z}, 
\\
h_i({\bf R}) &=& \frac{8 \pi}{3}| \Psi_i({\bf R})|^2 
\\
\nonumber
&& {} - \int d^3r |\Psi_i({\bf r})|^2 \frac{|{\bf r} - {\bf R}|^2
  - 3 [r_z - R_z]^2}{|{\bf r} - {\bf R}|^5},
\end{eqnarray}
written in atomic units; a factor of $\hbar/(4 \pi \epsilon_0)$ is
implied for the Hamiltonian.  The hyperfine interaction, $h_i({\bf r})$,
may be approximated by the dipolar interaction, $d({\bf r})$ when
${\bf r}$ is far outside the wave function of donor $i$.
Electron spin operators are written as
$\hat{S}$ with $i$ or $j$ indices and ${\bf R}_i$ position vectors.  Nuclear spin operators are written
as $\hat{I}$ with $n$ or $m$ indices and ${\bf r}_n$ position
vectors. 
The gyromagnetic ratios of the electron and $^{29}$Si nuclear spins
are $\gamma_E= 1.76 \times 10^{11} \mbox{(T s)$^{-1}$}$ and $\gamma_N =
5.31 \times 10^7 \mbox{(T s)$^{-1}$}$ respectively.  The wave function
of each donor electron, $\Psi_i({\bf r})$, is the Kohn-Luttinger wave function of a phosphorus donor impurity in
silicon, as described in Ref.~\cite{deSousaSD}.  In addition to the Hamiltonian-governed free evolution, we
model spin echo refocusing pulses as ideal spin flips.

\begin{figure}
\includegraphics[width=\linewidth]{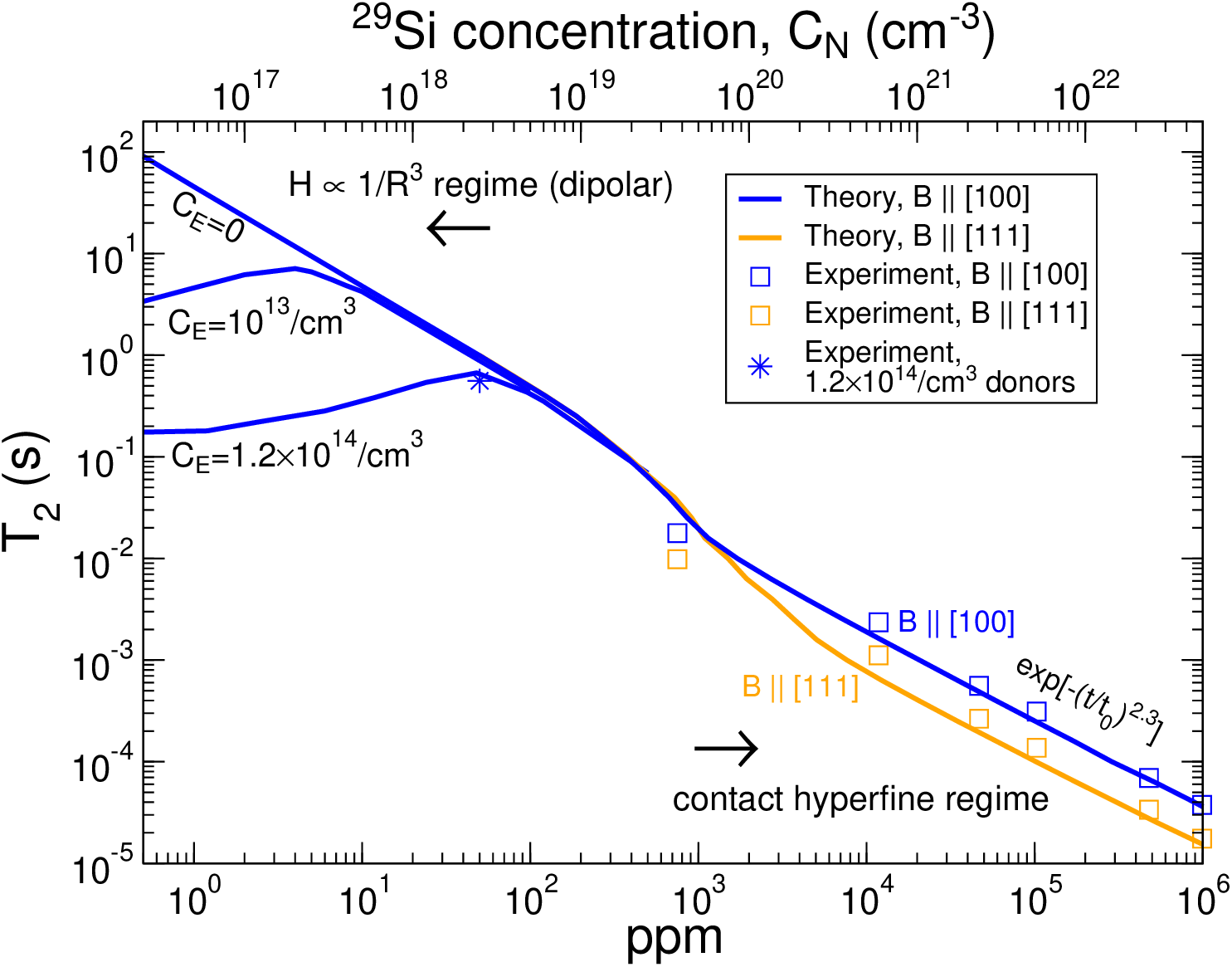}
\caption{
\label{T2vs29Si}
Decay times ($T_2$ for the Hahn echo) of Si:P donor electron spins
for various $C_N$. At high $C_N$, contact hyperfine interactions
dominate and $T_2$ is dependent upon the magnetic field direction
relative to the lattice orientation.  At low $C_N$, $T_2$ is
dependent upon $C_E$, and eventually dominated only by
dipolar interactions (which includes 
dipolar-approximated electronuclear interactions).
Experimental results are shown as square symbols, from Ref.~\onlinecite{Abe10}, and a
star symbol, from Ref.~\onlinecite{tyryshkin600ms}.
}
\end{figure}

To compute the decoherence time of a qubit in Si:P system, 
we take one of our donor electrons to be the
``central'' spin, say $i=0$, and simulate a Hahn spin echo on that
donor electron to remove the effects of static noise.  
Our dominant decoherence is due to
flip-flopping bath spins: $^{29}$Si-induced and donor-induced SD.
We display agreement with experiment for over $5$ orders of magnitude
in $C_{N}$, Fig.~\ref{T2vs29Si}, maintaining agreement
into very sparse densities.  This Letter presents procedures we have 
developed to accomplish this substantial (and very
computationally demanding) task.

We previously~\cite{WitzelSiP} computed the decoherence for
$C_N \gtrsim 1000~$ppm
using a cluster expansion technique which works well for dense spin baths.  
For sparse baths, we use the CCE~\cite{YangCCE}, applicable to both small
and large spin baths, with some adaptations.  
The CCE has a simple and self-evident
formulation which we now describe.  We define ${\cal L}(t) = \rho_{\uparrow \downarrow}(t)/\rho_{\uparrow \downarrow}(0)$, the off-diagonal
element of the reduced density matrix of our central spin after
performing a spin echo sequence over the duration $t = 2 \tau$, 
a refocusing pulse occurring at time $\tau$.  The
spin echo figure of merit is the modulus of ${\cal L}(t)$.  Next,
for a given set (cluster) of electron or nuclear bath spins, ${\cal
  C}$, we
define $L_{\cal C}(t)$ to be the resulting ${\cal L}(t)$ when we only include
flip-flop terms in our Hamiltonian [Eqs.~(\ref{Eq:H_E}-\ref{Eq:H_E-N})] that involve elements
of ${\cal C}$; {\it all} $\hat{S}_i^z \hat{S}_j^z$ interactions are
included in our implementation.  Then, we recursively define
\begin{equation}
\tilde{L}_{\cal C}(t) = L_{\cal C}(t) / \prod_{{\cal C}' \subset {\cal C}} \tilde{L}_{\cal C'}(t).
\end{equation}
By tautology, ${\cal L}(t) = \prod_{\cal C} \tilde{L}_{\cal C}(t)$,
providing a way to break the problem into independent factors coming
from each set of bath spins.  At short times, the smallest nontrivial
clusters dominate the decay; successively larger clusters become
significant with increasing evolution time.

\begin{figure}
\includegraphics[width=\linewidth]{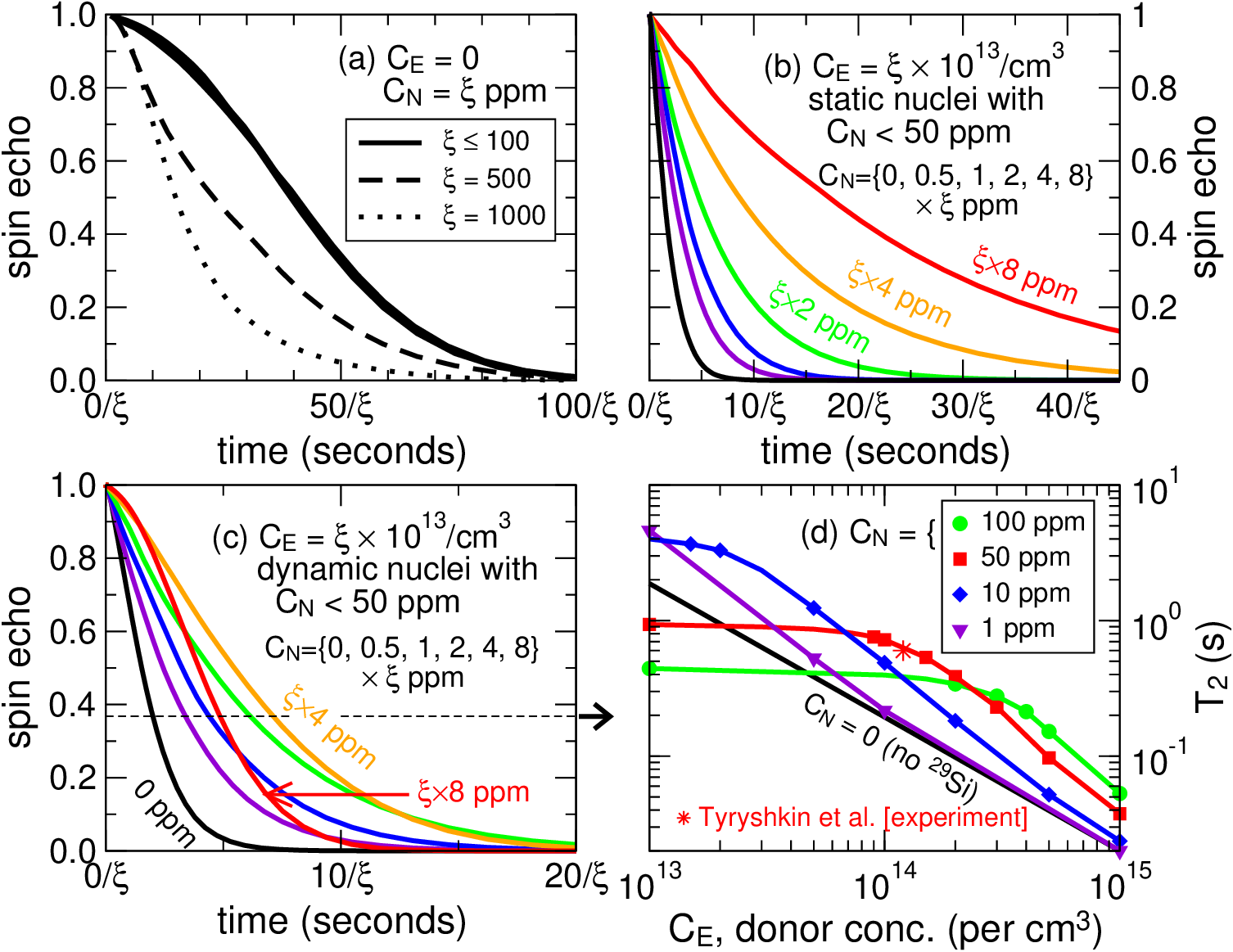}
\caption{
\label{quadT2}
Spin echo resulting from decoherence induced by various
concentrations of (a) $^{29}$Si ($C_N$), (b) background phosphorus
donors ($C_E$), or (c)
the combination of both.  
In (b), static $^{29}$Si-induced Overhauser field variations between donors
suppresses their decoherence-inducing flip-flops. 
(a)-(c) use $\xi$ as a scaling parameter for both concentration and
inverse time.
(d) $T_2$, defined as the $1/e$ decay time, for various $C_{N}$ and
$C_{E}$.  
The star symbol indicates an
experimentally obtained\cite{tyryshkin600ms} 
result at $C_N \approx 50~\mbox{ppm}$ and is in good agreement
with the corresponding $50~\mbox{ppm}$ theoretical curve. 
}
\end{figure}

These cluster expansions work well by
perturbative arguments in the
regime where the interactions among the bath spins are weak relative
to the interaction with the central spin.  Thus, $^{29}$Si-induced
SD is well-approximated when including only
2-clusters.  Donor-induced SD, however, is much more
challenging because the interaction strengths among the bath spins and
with the central spin are comparable.  In fact, we find that if we
compute the CCE expansion for different spatial configurations and
different initial spin states, the average over these
configurations and states can diverge rapidly when we include 4-clusters.  We
attribute this to the fact that different configurations of a sparse bath,
or even different states of a given spatial configuration,
can have very different convergence time scales for CCE.  However, we
find that the CCE is well behaved if we average over spin states within the
CCE definitions, that is, 
${\cal L}(t) = \langle \rho_{\uparrow \downarrow}^J(t) \rangle_J/\langle \rho_{\uparrow \downarrow}^J(0) \rangle_J$,
where $J$ represents each spin state and $\rho_{\uparrow \downarrow}^J$
is calculated with  $J$ as the initial bath state.

Of course, averaging over all spin states exactly would
be prohibitively difficult.  We find, however, that it is sufficient to
average over spin states in the following self-consistent
manner.
Choose a spatial configuration and a spin state
$\lvert J \rangle = \bigotimes_n \lvert j_n \rangle$
that serves as a template for spin state variants.  
Let $\Gamma$ be a set of clusters (e.g., up to a certain size) that
we include to approximate the solution: 
\begin{equation}
{\cal L}_{\Gamma}^J = \prod_{{\cal C} \in \Gamma}
\tilde{L}_{\cal C}^{{\cal K}(J, {\cal C}, \Gamma)},
\end{equation}
where ${\cal K}(J, {\cal C}, \Gamma)$ is the set of all spin
states that may differ from $J$ only for spins in
superclusters of ${\cal C}$ that are contained in $\Gamma$.
That is,
\begin{equation}
{\cal K}(J, {\cal C}, \Gamma) = \{J'|~\exists {\cal C}' \in \Gamma,
{\cal C}' \supseteq {\cal C}, {\cal D}(\lvert J \rangle, \lvert J'
\rangle) \subseteq {\cal C}'\},
\end{equation}
where ${\cal D}(\lvert J \rangle, \lvert J' \rangle)$ is the set of
spins whose state differs between $\lvert J \rangle$ and $\lvert J' \rangle$.
Then we define
\begin{equation}
\tilde{L}_{\cal C}^{\cal K} =
\langle L_{\cal C}^K \rangle_{K \in {\cal K}} / \prod_{{\cal C}'' \subset {\cal
    C}} \tilde{L}_{\cal C''}^{\cal K},
\end{equation}
where $L_{\cal C}^J$ solves the $L_{\cal C}$ problem for the given spin
state $J$.  Importantly, this yields the exact
spin state average solution for ${\cal L}_{\Gamma}^J$ in the
limit that $\Gamma$ includes all clusters ($J$ becomes
irrelevant).  Furthermore, it may be computed relatively efficiently.
With proper bookkeeping, each Hamiltonian 
(for a given cluster and external spin state) 
need only be diagonalized once, and
each $L_{\cal C}^J$ need only be computed once and raised to the
proper power to be multiplied into the solution.

\begin{figure}
\includegraphics[width=\linewidth]{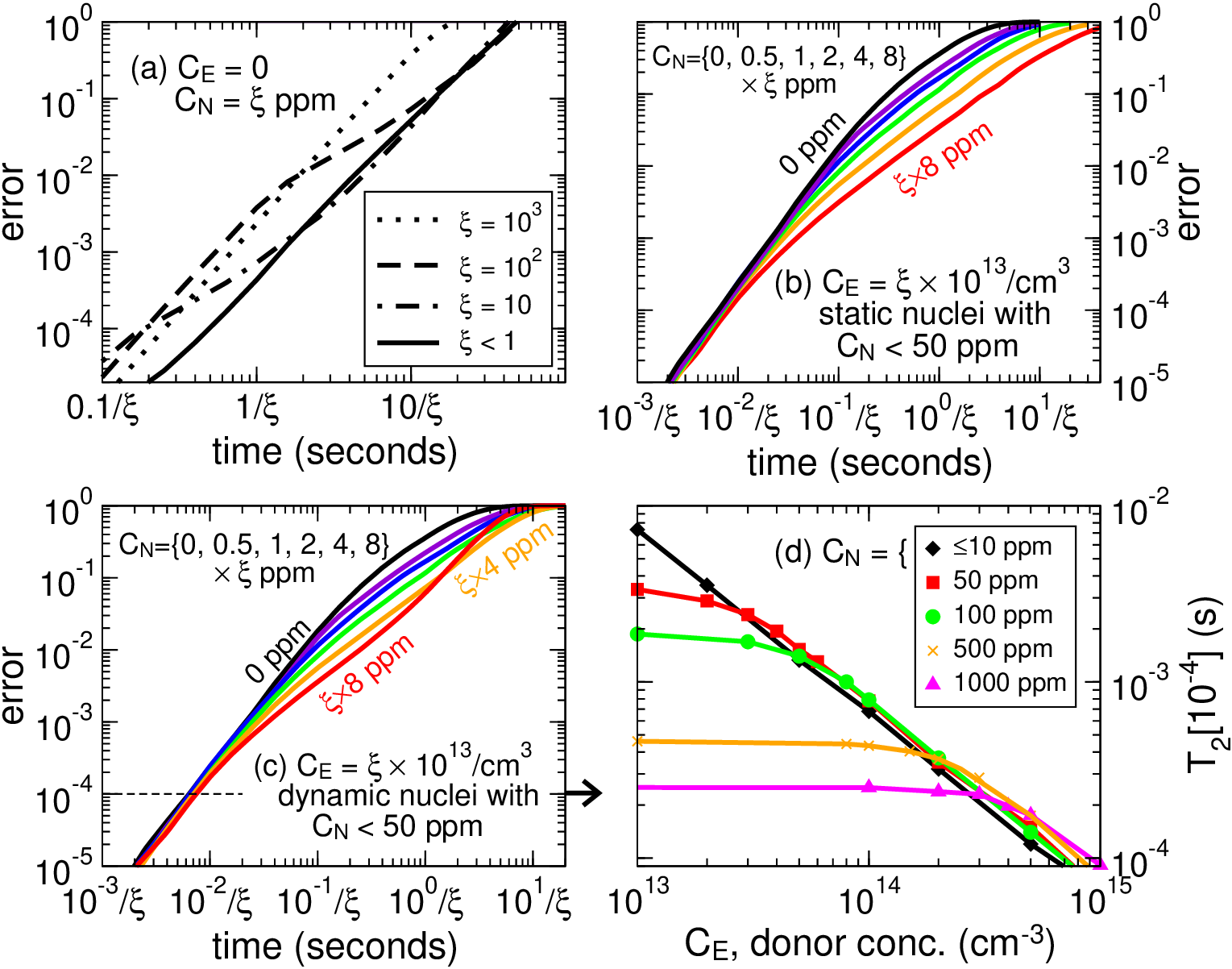}
\caption{
\label{quadTq}
(a)-(c) correlate with (a)-(c) of Fig.~\ref{quadT2},
respectively, but plotted in terms of error (one minus the spin echo)
on a log-log scale in order to highlight the low-error behavior that
is relevant for fault-tolerant quantum computation.  
(d) Plots $T_2[10^{-4}]$, the time at which the echo error is $10^{-4}$,
for various $C_N$ and $C_E$.
}
\end{figure}

We use heuristics and cutoffs to determine the $\Gamma$ 
set of clusters to include, trying to minimize the set necessary to
approximate the solution well.  We heuristically favor clusters with
strong interactions forming a connected graph over the entire
cluster and we employ cutoffs in the number of clusters, resonance
energies, and distance from the central spin.
We compute ensemble average results, such as shown in 
Figs.~\ref{T2vs29Si}, \ref{quadT2}, and \ref{quadTq}, 
by averaging results of different spatial configurations
and $J$ spin state templates for a given set of cutoffs.  These
cutoffs are adjusted until we obtain consistent, convergent results.

We present, in Figs.~\ref{quadT2} and \ref{quadTq}, ensemble averaged spin echo results for varied $C_E$ and $C_N$,
both separately and combined, 
We use $\xi$ as a scaling parameter to illustrate a perfect 
correspondence between concentrations and inverse time 
when decay is dominated entirely by $1/r^3$ dipolar interactions
($C_N \lesssim 50$~ppm).
These results show behavior ranging from decay dominated by
$^{29}$Si-induced SD to decay dominated by donor-induced SD.
When not dominated by $^{29}$Si-induced SD, the presence of $^{29}$Si
can actually prolong coherence, see Figs.~\ref{T2vs29Si} and \ref{quadT2}~(d), because
Overhauser field variations suppress donor flip-flops; 
a similar effect is noted~\cite{deSousaDonors} with respect to
qubit concentration and layout.
The initial decay, shown clearly in Fig.~\ref{quadTq}, 
behaves differently;
in this regime, any beneficial effect of $^{29}$Si is fairly insignificant.

\begin{figure}
\includegraphics[width=\linewidth]{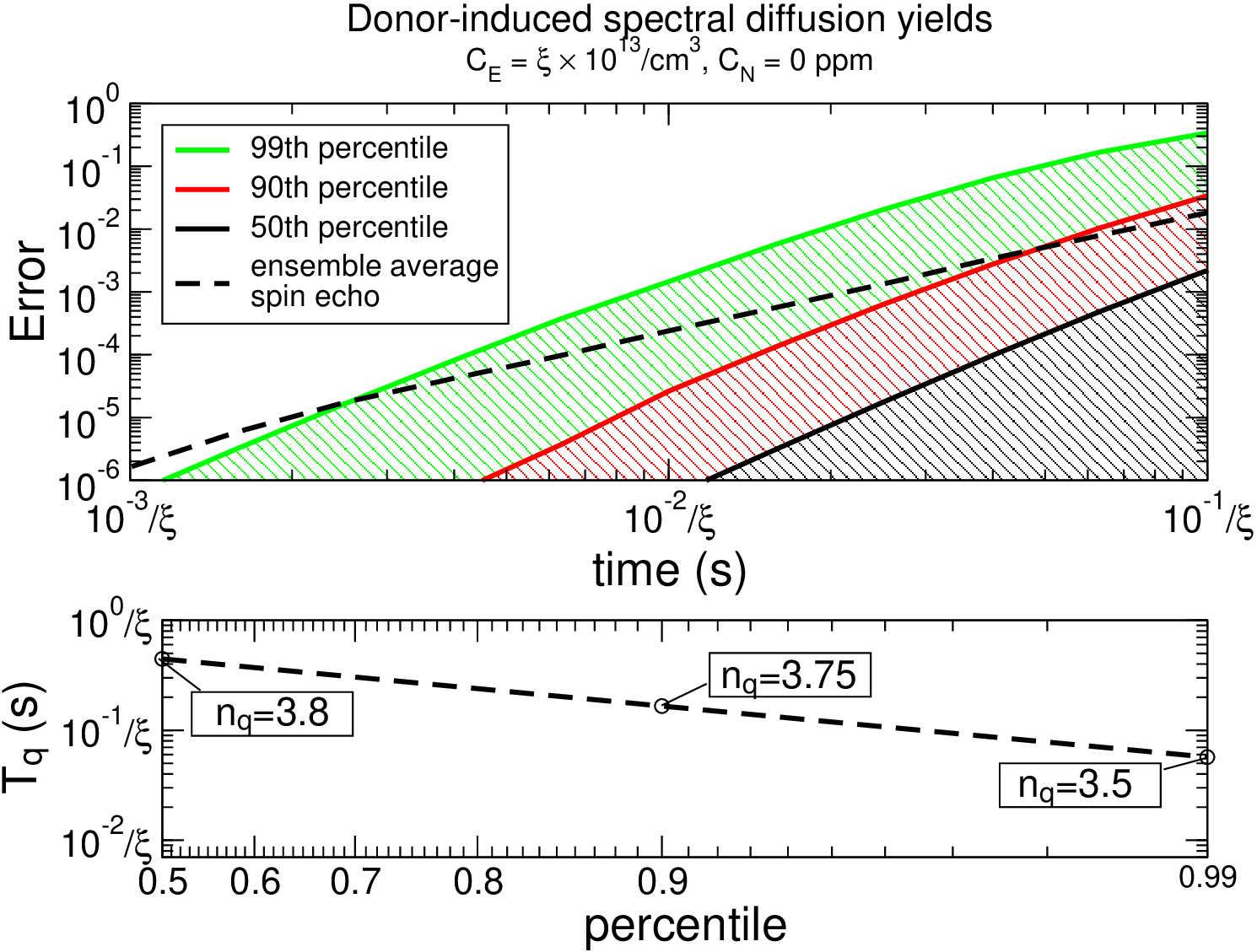}
\caption{
\label{yields}
(top) Maximum error from donor-induced (without $^{29}$Si)
SD for at least
$50\%$, $90\%$, or $99\%$ of the central spins due to random spatial 
variations of the bath spins.  (bottom) Corresponding fits of low-error behavior of
$1 - \exp{[-(t/T_q)^{n_q}]} \approx (t/T_q)^{n_q}$  in the $~10^{-4}$
error regime.
}
\end{figure}

Computing or measuring ensemble averages, as shown in
Figs.~\ref{quadT2} and \ref{quadTq}, has limited utility in the scope
of quantum computation.  It is more informative to compute the full
distribution of results that come out of the considerable sample-to-sample variation,
 which are significant especially for a central spin with dipolar
 coupling to a dilute bath~\cite{Dobrovitski_PRB08}.
Figure \ref{yields} addresses this for donor-induced spectral
diffusion 
by showing error distribution information for each spin echo time
independently; essentially, this gives a performance guarantee for
various fractions of possible donors.  
At short times, the ensemble average echo decay error is actually 
dominated by statistical 
outliers as the top panel of Fig.~\ref{yields} demonstrates.
Used in the bottom panel of Fig.~\ref{yields}, 
we introduce $T_q$ and $n_q$ as figures of merit 
that characterize initial decoherence at short
times, as appropriate for quantum information considerations.  These 
are obtained by fitting the error to $1 - \exp{[-(t/T_q)^{n_q}]}
\approx (t/T_q)^{n_q}$ in the $~10^{-4}$
error regime (motivated by common fault-tolerance thresholds).
These results have direct implications for quantum computer architecture
designs and error analysis~\cite{QCarchitecture}.

Apart from reducing $C_E$, donor-induced SD may be
suppressed by polarizing the background donors thermally at temperatures
that are readily achieved in specialized refrigerators~\cite{Morello06}.
In the low-error limit where donor-induced SD often
dominates [Fig.~\ref{quadTq}], the error is proportional to the number
of contributing 2-clusters which must have opposite spin polarization in order to
flip-flop; thus, $T_q \propto (p_{\uparrow} p_{\downarrow})^{-n_q}$,
where $p_{\uparrow / \downarrow} = \exp{(\pm E_z / k_B T)} / 2 \cosh{(E_z / k_B T)}$ from
Boltzmann statistics with $E_Z$ as the electron Zeeman energy
splitting corresponding to about $1.3$ K per Tesla.

To conclude, we adapted the cluster correlation
expansion~\cite{YangCCE}, by retaining all Ising-like interactions and
interlacing spin state averaging in a self-consistent manner, 
to study decoherence induced by a background of dynamical donor 
electron spins in silicon.  We demonstrate
that approaching the $T_2 =
2 T_1$ limit through isotopic enrichment in Si is 
impossible in the presence of a finite concentration of unpolarized donors.  
Unavoidable donor impurities in the background
make this limit, where $T_1$ of one hour has been reported at
$1.25~$K~\cite{1hourT1}, unattainable, though prospects
improve if the electrons may be thermally polarized.
While the presence of some $^{29}$Si can actually
increase $T_2$ considerably by suppressing donor-induced decoherence, 
this effect is fairly insignificant in the short time (low-error)
regime important for quantum computation.
We introduce $T_q$ and $n_q$ to describe decoherence at short times 
and discuss the effect of statistical variation of impurity locations 
on decoherence.  Variation in the decoherence of different donors
becomes extremely significant in the regime of low impurity 
concentration, a crucial consideration for designing 
quantum computer architectures and determining fabrication requirements.

We thank A. Tyryshkin, S. Lyon, C. Tahan, R. Muller, E. Nielsen, R. Rahman, A. Ganti, and
A. Landahl for comments.  
Sandia National Laboratories is a
multiprogram laboratory operated by Sandia Corporation, a wholly
owned subsidiary of Lockheed Martin Corporation, for the U.S. Department
of Energy's National Nuclear Security Administration under contract
DE-AC04-94AL85000.
SDS and {\L}C acknowledge LPS-NSA support; 
{\L}C is also supported by the Homing programme of the
Foundation for Polish Science and the EEA Financial
Mechanism.  
AM is supported by the Australian Research Council, Australian
Government, U.S. NSA, and U.S. ARO under contract No. W911NF-08-1-0527.

\end{document}